\documentclass[11pt]{article}
\usepackage{amssymb}
\usepackage{amsmath,url,hyperref,fullpage,graphicx}

\newtheorem{Theorem}{Theorem}
\newtheorem{Property}{Property}
\newenvironment{proof}{\paragraph{Proof:}}{\hfill$\square$}

\def\Holder{\mathrm{H\ddot{o}lder}}

\def\eqdef{:=}
\def\st{\ :\ }
\def\calM{\mathcal{M}}
\def\calF{\mathcal{F}}
\def\calX{\mathcal{X}}
\def\calN{\mathcal{N}}
\def\calE{\mathcal{E}}
\def\dmu{\mathrm{d}\mu}
\def\CS{\mathrm{CS}}
\def\bbR{\mathbb{R}}

\title{A note on Onicescu's informational energy and correlation coefficient in exponential families}
 
\author{Frank Nielsen\footnote{E-mail: {\tt Frank.Nielsen@acm.org}}\\ Sony Computer Science Laboratories Inc.\\ Tokyo, Japan}

\date{March 2020}

\begin{document}
\maketitle

\begin{abstract}
The informational energy of Onicescu is a positive quantity that measures 
the amount of uncertainty of a random variable. 
But contrary to Shannon's entropy, the informational energy increases when randomness decreases.
We report closed-form formula for Onicescu's informational energy and its associated correlation coefficient when the probability distributions belong to an exponential family. We show how to instantiate the generic formula for several common exponential families. 
\end{abstract}

\noindent{Keywords}: Shannon's entropy, Onicescu's informational energy, Onicescu's correlation coefficient, exponential families, Cauchy-Schwarz divergence.

\section{Introduction}

\subsection{Onicescu's informational energy}

Let $(\calX,\calF,\mu)$ be a probability space~\cite{PM-1995} with $\sigma$-algebra $\calF$ on the sample space $\calX$, and $\mu$ a base  measure often chosen as the Lebesgue measure or as the counting measure. 
Let $\calM$ denote the set of Radon-Nikodym densities of probability measures dominated by $\mu$.
Two probability densities $p$ and $q$ are said equal (i.e., $p=q$) if and only if $p(x)=q(x)$ $\mu$-almost everywhere, and different (i.e., $p\not=q$) 
when $\mu\left(\{x\in\calX\st p(x)=q(x)\}\right)\not =0$.

Octav Onicescu~\cite{Ob1986,crepel2013statisticians} (1892-1983) was a renowned Romanian mathematician who founded the school of probability theory and statistics~\cite{onicescu1979elements} in Romania.
Onicescu introduced the  informational energy~\cite{Onicescu-1966} (also termed information energy~\cite{pardo1991information} in the literature) of a probability measure $P\ll\mu$ with Radon-Nikodym density $p=\frac{\mathrm{d}P}{\mathrm{d}\mu}$ as
\begin{equation}
I(p) \eqdef \int p^2(x) \dmu(x) >0.
\end{equation}

The R\'enyi entropy~\cite{pardo1994asymptotic,nielsen2011r} of order-$2$ can be written using the informational energy:
\begin{equation}
R_2(p) \eqdef  -\log \int p^2(x) \dmu(x) = -\log I(p),
\end{equation}
as well as Vajda's quadratic entropy~\cite{vajda2007generalized}:
\begin{equation}
V_2(p) \eqdef  1-  \int p^2(x) \dmu(x) =1-I(p).
\end{equation}

Notice that it follows from Cauchy-Schwarz's inequality that we have $I(p)\leq 1$ when $p\in L^2(\mu)$, the Lebesgue space  of square integrable functions~\cite{PM-1995}.
The informational energy for continuous distributions $p$  is $I(p) = \int_\calX p^2(x) \mathrm{d}x$ and the informational energy   for discrete distributions $q$
 $I(q)=\sum_{x\in\calX} q^2(x)$. 
Notice that the informational energy in the continuous case is
not a limit of the informational energy in the discrete case~\cite{pardo1991information,ho2009discontinuity}.

The informational energy is an important concept in statistics  which fruitfully interplays with Shannon's entropy~\cite{cover2012elements} $H(p)$:
\begin{equation}
H(p) \eqdef  \int -p(x)\log p(x) \dmu(x).
\end{equation}

Notice that the informational energy is always positive but the Shannon's entropy may be negative for continuous distributions (e.g., differential entropy of the normal distributions for small standard deviation).
For the Dirac's distribution  $\delta_{e}$ (with $\delta_{e}(x)=1$ when $x=e$ and $0$ otherwise), the informational energy is 
$I(\delta_{e})=+\infty$ but Shannon's entropy is $H(\delta_{e})=-\infty$.

The informational energy measures the amount of uncertainty of a random variable like Shannon's entropy but augments when randomness decreases.
The informational energy was originally motivated by an analogy to the kinetic energy in physics, and  proves useful when investigating the thermodynamics laws on a statistical manifold~\cite{IG-CalinUdriste-2014} where thermodynamic processes are viewed as trajectories on the manifold.
Another key difference with Shannon's entropy is that Shannon's entropy is always strictly concave~\cite{cover2012elements} but the informational energy is always strictly convex:

\begin{Property}
Onicescu's informational energy $I(\cdot)$ is a strictly convex functional.
\end{Property}

\begin{proof}
A function $F$ is strictly convex iff for any $\alpha\in(0,1)$ and $p\not =q$ two densities $\calM$, we have
$F((1-\alpha)p+\alpha q)< (1-\alpha)F(p)+\alpha F(q)$.
Let us check this strict inequality for the informational energy $I$:
\begin{eqnarray}
I((1-\alpha)p+\alpha q) &=& (1-\alpha)^2I(p)+\alpha^2I(q)+2\alpha(1-\alpha)\int p(x)q(x)\dmu(x),\\
&=& (1-\alpha)I(p)+\alpha I(q)+2\alpha(1-\alpha)\int p(x)q(x)\dmu(x)-\alpha(1-\alpha)(I(p)+I(q)),\nonumber\\
&=&  (1-\alpha)I(p)+\alpha I(q) - \underbrace{\alpha(1-\alpha) \int (p(x)-q(x))^2\dmu(x)}_{>0},\\
&<&  (1-\alpha)I(p)+\alpha I(q),
\end{eqnarray}
when $p\not=q$ and $\alpha\in(0,1)$.
\end{proof}

Since $I(p)$ is strictly convex, we can define the informational energy divergence as the following Jensen divergence~\cite{nielsen2011burbea} measuring the convexity gap:
\begin{eqnarray}
J_I(p,q) &\eqdef& \frac{I(p)+I(q)}{2}-I\left(\frac{p+q}{2}\right),\\
&=& \frac{1}{4}\int_\calX (p(x)-q(x))^2\dmu(x).
\end{eqnarray}

For a uniform discrete distribution $p$ on an alphabet $\calX$ of $d$ letters, we have $I(p)=\frac{1}{d}$, 
and for any probability mass function $p$ on $\calX$, we have $I(p)\geq \frac{1}{d}$.
Recall that $H(p)\leq\log d$ with equality when $p$ is the discrete uniform distribution. 
More generally, for a continuous density on an interval $\calX=[a,b]$, we have $I(p)\geq \frac{1}{b-a}$.

For discrete or continuous distributions, we have the following inequality~\cite{IG-CalinUdriste-2014} (Proposition 5.8.5):
\begin{equation}
H(p)+\frac{1}{2}I(p)\geq 1-\log 2 > 0.69897,
\end{equation}
and Shannon's cross-entropy
\begin{equation}
H^\times(p:q) =  \int -p(x)\log q(x) \dmu(x)
\end{equation}
can be lower bounded using the informational energy as follows (Problem 5.8 in~\cite{IG-CalinUdriste-2014}):
For any $x>0$, we have $\log x\leq x-1$.
Thus we get $\log q(x)\leq q(x)-1$ and $-p(x)\log q(x)\geq p(x)-p(x)q(x)$.
Therefore we have
$$
H^\times(p:q) \geq 1-\int p(x)q(x) \dmu(x).
$$

Using Cauchy–Schwarz's inequality $\int p(x)q(x)\dmu(x)\leq \sqrt{\int p(x)^2\dmu(x)}\, \sqrt{\int q(x)^2\dmu(x)}$, we get:

\begin{equation}
H^\times(p:q) \geq 1-\sqrt{I(p)\, I(q)}.
\end{equation}

In particular, we get a lower bound on Shannon's entropy: 
$$
H(p)=H^\times(p:q)\geq 1-I(p).
$$

Table~\ref{tab:comparison} summarizes the comparison between Shannon's entropy and Onicescu's informational energy:

\begin{table}
\centering
\begin{tabular}{lll}
 & Entropy $H(p)$ & Informational energy $I(p)$\\ \hline
convexity & strictly concave & strictly convex \\
range & can be negative & always positive \\
uncertainty measure & augments with disorder & decreases with disorder \\
uniform discrete distribution $u$  & $H(u)=\log d$ & $I(u)=\frac{1}{d}$\\
(with alphabet size $|\calX|=d$) & \\
bound & $H(p)\geq 1-I(p)$  & $I(p)\geq 1-H(p)$\\
\multicolumn{3}{c}{Inequality: $H(p)+\frac{1}{2}\, I(p)\geq 1-\log 2$}
\end{tabular}
\caption{Comparison between Shannon's entropy and Onicescu's informational energy.\label{tab:comparison}}
\end{table}

For an in-depth treatment of Onicescu's informational energy, we refer to the paper~\cite{pardo1991information} ($77$ pages, with main properties listed in pp. 167-169 as statistical applications).
Onicescu's informational energy has been used in physics~\cite{agop2015implications}, information theory in electronic structure theory of atomic and molecular systems~\cite{chatzisavvas2005information,alipour2012onicescu,ou2019shannon}, machine learning~\cite{andonie2004information}, and complex systems~\cite{rizescu2014using} among others.

\subsection{Onicescu's correlation coefficient}

Onicescu also defined a correlation coefficient (see~\cite{IG-CalinUdriste-2014}, Chapter~5):
\begin{equation}\label{eq:rho}
\rho(p,q) \eqdef \frac{I(p,q)}{\sqrt{I(p)\, I(q)}},
\end{equation}
where $I(p,q)$ denotes the cross-informational energy:
\begin{equation}
I(p,q) \eqdef \int p(x)q(x) \dmu(x),
\end{equation}
with $I(p)=I(p,p)$.
Notice that it follows from the Cauchy-Schwarz inequality that $I(p,q)\leq \sqrt{I(p)I(q)}$, and therefore we have: 
$$
0<\rho(p,q)\leq 1,
$$
assuming both densities $p$ and $q$ belong to the Lebesgue space $L^2(\mu)$.

Notice that the informational energy of a statistical mixture $m(x)=\sum_{i=1}^k w_ip_i(x)$ with $k$ weighted components $p_1(x),\ldots, p_k(x)$ (with $w\in\Delta_{k}$ the $(k-1)$-dimensional standard simplex) can be expressed as follows:
\begin{equation}
I(m)= \int \left(\sum_{i=1}^k w_ip_i(x)\right)^2 \dmu(x) = \sum_{i=1}^k \sum_{j=1}^k w_iw_j\, I(p_i,p_j).
\end{equation} 

The Cauchy-Schwarz divergence~\cite{jenssen2006cauchy,nielsen2017holder} is defined by
\begin{equation}
D_\CS(p,q) \eqdef -\log\left( \frac{\int_{\calX} p(x) q(x) \dmu(x)}{\sqrt{\left(\int_{\calX} p(x)^{2} \dmu(x)\right)\left(\int_{\calX} q(x)^{2}\dmu(x)\right)}} \right) \geq 0.
\end{equation}
Thus the Cauchy-Schwarz divergence is a projective divergence (that is, we have $D_\CS(p,q) =D_\CS(\lambda p,\lambda' q)$ for any $\lambda>0$ and $\lambda'>0$) which can be rewritten using the Onicescu's correlation coefficient as:
\begin{equation}
D_\CS(p,q)= -\log\left(\rho(p,q)\right).
\end{equation}

\subsection{Exponential families}
Consider a natural exponential family~\cite{nielsen2009statistical,Barndorff-2014} (NEF)
\begin{equation}
\calE= \left\{p_\theta(x)=\exp\left(\theta^\top t(x)-F(\theta)+k(x)\right) \ : \ \theta\in\Theta\right\},  \label{eq:factorization}
\end{equation}
where $t(x)$ denotes the minimal sufficient statistics, $k(x)$ an auxiliary measure carrier term, and 
\begin{equation}
F(\theta) := \log\left(\int_\calX \exp(\theta^\top t(x)) \dmu(x)\right),
\end{equation}
the cumulant function which is commonly called the log-normalizer (or log-partition function in statistical physics).
Parameter $\theta$ is called the natural parameter and is defined on the open convex natural parameter space $\Theta$.
 
Many familiar families of distributions $\{p_\lambda(x)\ \lambda\in\Lambda\}$ are exponential families in disguise after reparameterization: 
$p_\lambda(x)=p_{\theta(\lambda)}(x)$ (e.g., normal family or Poisson family). 
Those families are called exponential families (omitting the leading adjective `natural'), and their densities are canonically factorized as follows:
\begin{equation}\label{eq:canef}
p_\lambda(x)=\exp\left(\theta(\lambda)^\top t(x)-F(\theta(\lambda))+k(x)\right).
\end{equation}
We  call parameter $\lambda\in\Lambda$ the  source parameter, and parameter $\theta(\lambda)\in\Theta$ is called the corresponding natural parameter.
Densities of an exponential family have all the same support $\calX$.

\section{Onicescu's informational energy and correlation coefficient in exponential families}

We report closed-form formulas for Onicescu's informational energy and correlation coefficient when densities belong to a prescribed exponential family, and then illustrate those formula on common families of probability distributions.

\begin{Theorem}[Onicescu's informational energy and correlation coefficient in exponential families]
In an exponential family $\calE=\{p_\theta\}_{\theta\in\Theta}$, Onicescu's informational energy of a probability density $p_\theta$ is given by:
\begin{equation}\label{eq:icf}
I(p_\theta) =  
 \exp\left(F(2\theta)-2F(\theta)\right)\, E_{p_{2\theta}}\left[\exp(k(x))\right],
\end{equation}
provided that $2\theta\in\Theta$ so that $p_{2\theta}\in\calE$. 
When the auxiliary carrier term $k(x)$ vanishes, we have $E_{p_{2\theta}}\left[\exp(k(x))\right]=1$.

The Onicescu's correlation coefficient $\rho(p_{\theta_1},p_{\theta_2})$ between densities $p_1=p_{\theta_1}$ and $p_2=p_{\theta_2}$ is
\begin{equation}\label{eq:rhocf}
\rho(p_{\theta_1},p_{\theta_2}) = \exp(-J_F(2\theta_1:2\theta_2)) \times 
\frac{E_{p_{\theta_1+\theta_2}}\left[\exp(k(x))\right]}{\sqrt{E_{p_{2\theta_1}}\left[\exp(k(x))\right] \, E_{p_{2\theta_2}}\left[\exp(k(x))\right]}},
\end{equation}
provided that $\theta_1+\theta_2\in\Theta$, $2\theta_1\in\Theta$, $2\theta_2\in\Theta$,
where 
\begin{equation}
J_F(\theta_1,\theta_2)\eqdef\frac{F(\theta_1)+F(\theta_2)}{2}-F\left(\frac{\theta_1+\theta_2}{2}\right)\geq 0,
\end{equation}
is a Jensen divergence~\cite{nielsen2011burbea} induced by the cumulant function of the exponential family.
\end{Theorem}

\begin{proof}
The proof follows the same line of arguments as in~\cite{nielsen2011closed}.
Consider the term $I(p_{\theta_1},p_{\theta_2})$:
\begin{eqnarray}
I(p_{\theta_1},p_{\theta_2})&=& \int \exp(t(x)^\top\theta_1-F(\theta_1)+k(x))  \exp(t(x)^\top\theta_2-F(\theta_2)+k(x))\dmu(x),\\
&=&  \int \exp\left(t(x)^\top (\theta_1+\theta_2)-F(\theta_1+\theta_2)+k(x)+F(\theta_1+\theta_2)-F(\theta_1)-F(\theta_2)+k(x)\right) \dmu(x),\nonumber\\
&=& \exp\left(F(\theta_1+\theta_2)-F(\theta_1)-F(\theta_2)\right)\ \int p_{\theta_1+\theta_2}(x) \exp(k(x))\dmu(x),\\
&=&  \exp\left(F(\theta_1+\theta_2)-F(\theta_1)-F(\theta_2)\right)\ E_{p_{\theta_1+\theta_2}}\left[\exp(k(x))\right],
\end{eqnarray}
provided that $\theta_1+\theta_2\in\Theta$.
This condition is always satisfied when the natural parameter space 
is either a cone~\cite{nielsen2011closed} (e.g., Gaussian family, Wishart family, etc) or an affine space~\cite{nielsen2013chi} (e.g., Poisson family, isotropic Gaussian family, etc).
Since $I(p)=I(p,p)$ and $\rho(p,q)=\frac{I(p,q)}{\sqrt{I(p)I(q)}}$, we deduce formula of Eq.~\ref{eq:icf} and Eq.~\ref{eq:rhocf}.
\end{proof}

Since $D_\CS(p,q)= -\log\left(\rho(p,q)\right)$, we get the following closed-form for the Cauchy-Schwarz divergence:
\begin{equation}
D_\CS(p_{\theta_1},p_{\theta_2})=J_F(2\theta_1:2\theta_2)+\log\left( \frac{\sqrt{E_{p_{2\theta_1}}\left[\exp(k(x))\right] \, E_{p_{2\theta_2}}\left[\exp(k(x))\right]}}{E_{p_{\theta_1+\theta_2}}\left[\exp(k(x))\right]}\right).
\end{equation}
We check that when $\theta_1=\theta_2$, we have $D_\CS(p_{\theta_1},p_{\theta_2})=0$.
Closed-form formula were also reported for the Cauchy-Schwarz divergence between densities of an exponential family in~\cite{nielsen2012closed}.
Table~\ref{tab:comp} reports the formula for Onicescu's informational energy and the Shannon's entropy~\cite{nielsen2010entropies} for densities belonging to some common exponential families.
These formula can be recovered easily from the generic formula using the canonical decompositions of exponential families 
reported in~\cite{nielsen2009statistical}.
Shannon's entropy~\cite{nielsen2010entropies} of a density $p_\theta$ of $\calE$ is
\begin{eqnarray}
H(p_\theta)&=& F(\theta)-\theta^\top\nabla F(\theta)-E_{p_\theta}[k(x)],\\ 
H(p_\theta)&=& -F^*(\eta)-E_{p_\theta}[k(x)], 
\end{eqnarray}
where $F^*$ denotes the Legendre-Fenchel convex conjugate and $\eta=\nabla F(\theta)=E_{p_\theta}[t(x)]$ the moment parameter. 

Notice that when $k(x)=0$ (no auxiliary carrier term, e.g., Gaussian family), we have $E_p[e^{k(x)}]=E_p[1]=\int p(x)\dmu(x)=1$ for any density $p\in\calM$.
In that case, the above formula simplify as follows:
\begin{eqnarray}
I(p_\theta) &=& 
\exp\left(F(2\theta)-2F(\theta)\right),\\
\rho(p_{\theta_1},p_{\theta_2}) &=& \exp(-J_F(2\theta_1:2\theta_2)),\\
D_\CS(p_{\theta_1},p_{\theta_2})&=& J_F(2\theta_1:2\theta_2).
\end{eqnarray}

The Cauchy-Schwarz divergence between mixtures of Gaussians has been reported in~\cite{kampa2011closed}, and extended to mixtures of exponential families with conic natural parameter spaces in~\cite{nielsen2012closed}.

Furthermore, since the Jensen divergence $J_F$ is defined for a strictly convex generator $F$ modulo an affine term, we may choose
the representative $F(\theta)=-\log p_\theta(\omega)=:-l_\theta(\omega)$ for the equivalence class $[F]$ of strictly convex functions, where $\omega$ is any point belonging to the the support $\calX$ of the exponential family $\calX$ and $l_\theta(\cdot)$ the (concave) log-likelihood function, see~\cite{nielsen2020cumulant} for details.

It follows that, we can rewrite the Onicescu's informational energy, correlation coefficient and the Cauchy-Schwarz divergence when $k(x)=0$ as follows:
\begin{eqnarray}
I(p_\theta) &=&  \frac{p_\theta^2(\omega)}{p_{2\theta}(\omega)}, \quad\forall\omega\in\calX \label{eq:iecfk}\\
\rho(p_{\theta_1},p_{\theta_2}) &=&\frac{\sqrt{p_{2\theta_1}(\omega)}\sqrt{p_{2\theta_2}(\omega)}}{p_{\theta_1+\theta_2}(\omega)},\quad\forall\omega\in\calX,\label{eq:rhoomega}\\
D_\CS(p_{\theta_1},p_{\theta_2})&=& -\log\left(  \frac{\sqrt{p_{2\theta_1}(\omega)}\sqrt{p_{2\theta_2}(\omega)}}{p_{\theta_1+\theta_2}(\omega)} \right) ,\\
 &=& l_{\theta_1+\theta_2}(\omega)-\frac{l_{2\theta_1}(\omega)+ l_{2\theta_2}(\omega)}{2},\quad\forall\omega\in\calX.
\end{eqnarray}

Moreover, the Cauchy-Schwarz divergence can be generalized to the broader class of H\"older divergences~\cite{nielsen2017holder} for conjugate exponents $\frac{1}{\alpha}+\frac{1}{\beta}=1$ and $\alpha>0$ and $\gamma>0$ as follows:
\begin{eqnarray}
D_\Holder^{\alpha,\gamma}(p_{\theta_1},p_{\theta_2})&\eqdef&  
-\log \left(\frac{\int_{\mathcal{X}} p(x)^{\gamma / \alpha} q(x)^{\gamma / \beta} \dmu(x)}{\left(\int_{\mathcal{X}}
 p(x)^{\gamma} \dmu(x)\right)^{1 / \alpha}\left(\int_{\mathcal{X}} q(x)^{\gamma} \dmu(x)\right)^{1 / \beta}}\right)
,\\
&=& \log\left( 
\frac{p_{\frac{\gamma}{\alpha}\theta_1+\frac{\gamma}{\beta}\theta_2}(\omega)}{p_{\gamma\theta_1}^{\frac{1}{\alpha}}(\omega)
p_{\gamma\theta_2}^{\frac{1}{\beta}}(\omega)  }\right), \quad\forall\omega\in\calX,\\
D_\Holder^{\alpha,\gamma}(p_{\lambda_1},p_{\lambda_2}) &=& 
\log\left( 
\frac{p_{\frac{\gamma}{\alpha}\theta(\lambda_1)+\frac{\gamma}{\beta}\theta(\lambda_2)}(\omega)}{p_{\gamma\theta(\lambda_1)}^{\frac{1}{\alpha}}(\omega)
p_{\gamma\theta(\lambda_2)}^{\frac{1}{\beta}}(\omega)  }\right), \quad\forall\omega\in\calX.
\end{eqnarray}
The latter two formula hold when $k(x)=0$ (no auxiliary carrier term like the Gaussian family).
When $\alpha=\beta=\gamma=2$, we recover the Cauchy-Schwarz divergence: 
$D_\Holder^{2,2}(p_{\theta_1},p_{\theta_2})=D_\CS(p_{\theta_1},p_{\theta_2})$.

\begin{table}
\centering
$$
\begin{array}{l|ll}
\mbox{Family} & \mbox{entropy} & \mbox{informational energy}\\ \hline
\mbox{Generic $\calE$} & F(\theta)-\theta^\top\nabla F(\theta)-E_{p_\theta}[k(x)] & e^{F(2\theta)-2F(\theta)} E_{p_{2\theta}}\left[\exp(k(x))\right] \\ \hline
\mbox{Normal $N(\mu,\sigma)$}  &  \frac{1}{2}\log(2\pi e\sigma^2) & \frac{1}{2\sigma\sqrt{\pi}}\\
\mbox{Multivariate Normal $N(\mu,\Sigma)$} & \frac{1}{2} \log |2\pi e\Sigma| & \pi^{-\frac{d}{2}}2^{-d}|\Sigma|^{-\frac{1}{2}} \\ 
\mbox{$\mathrm{LogNormal}(\mu,\sigma)$}  & \log(\sigma e^{\mu+\frac{1}{2}}\sqrt{2\pi}) &  \frac{1}{2\sigma\sqrt{\pi}}\exp(\frac{\sigma^2}{4}-\mu) \\
\mbox{$\mathrm{Exponential}(\lambda)$} & 1-\log\lambda & \frac{\lambda}{2}\\
\mbox{$\mathrm{Pareto}_k(a)$} & 1+\frac{1}{a}+\log\frac{k}{a} &  \frac{a^2}{k(2a+1)}\\
\mbox{$\mathrm{Gamma}(\alpha,\beta)$} & \alpha+\log\frac{\Gamma(\alpha)}{\beta}+(1-\alpha)\psi(\alpha) & \frac{1}{\beta(2\alpha-1)B\left(\alpha,\frac{1}{2}\right)}\\
\mbox{$\mathrm{Beta}(\alpha,\beta)$} &
\log \mathrm{B}(\alpha, \beta)-(\alpha-1) \psi(\alpha)
 & B^2(\alpha,\beta)\frac{\Gamma(2\alpha-1)\Gamma(2\beta-1)}{\Gamma(2\alpha+2\beta-2)} \\
& -(\beta-1) \psi(\beta)  & \\
& +(\alpha+\beta-2) \psi(\alpha+\beta) & \\
\mbox{$\mathrm{Poisson}(\lambda)$} & \lambda(1-\log\lambda) +e^{-\lambda}\sum_{i=0}^\infty \frac{\lambda^i\log i!}{i!} & 
\exp(-2\lambda) \sum_{i=0}^\infty \frac{\lambda^{2i}}{(i!)^2}
\end{array}
$$

\caption{Comparisons between Shannon's entropy and Onicescu's informational energy for common distributions of exponential families.
 $|\cdot|$ denotes the matrix determinant, $\Gamma(\cdot)$  the gamma function, $\psi(\cdot)$ the digamma function, and $\mathrm{B}(\alpha, \beta)=\frac{\Gamma(\alpha) \Gamma(\beta)}{\Gamma(\alpha+\beta)}$.
\label{tab:comp}}
\end{table}

\section{Some illustrating examples}

Let us illustrate how to instantiate the generic formula with some examples of exponential families.

\subsection{Exponential family of exponential distributions}

Consider the family of exponential distributions with rate parameter $\lambda>0$.
The densities of this exponential family writes as $p_\lambda(x)=\lambda\exp(-\lambda x)$ with  support $\calX=[0,\infty)$.
We use the canonical decomposition of the exponential family to get $t(x)=-x$, $\theta=\lambda$, $F(\theta)=-\log\theta$ and $k(x)=0$.
It follows that
\begin{eqnarray}
I(p_\theta) &=& \exp\left(F(2\theta)-2F(\theta)\right),\\
 &=& \exp\left(-\log 2\theta +2\log\theta\right),\\
&=&  \exp\left(-\log 2+\log\theta\right),\\
&=& \frac{\theta}{2}.
\end{eqnarray}
Thus $I(p_\lambda)=\frac{\lambda}{2}$.
Similarly, we find that 
\begin{eqnarray}
I(p_{\lambda_1},p_{\lambda_2})&=& \exp \left(F\left(\theta_{1}+\theta_{2}\right)-F\left(\theta_{1}\right)-F\left(\theta_{2}\right)\right) E_{p_{\theta_{1}+\theta_{2}}}[\exp (k(x))],\\
&=& \frac{\lambda_1\lambda_2}{\lambda_1+\lambda_2}.
\end{eqnarray}
Thus $\rho(p_{\lambda_1},p_{\lambda_2})=\frac{2\sqrt{\lambda_1\lambda_2}}{\lambda_1+\lambda_2}$,
and $D_\CS(p_{\lambda_1},p_{\lambda_2})=\log\left(\frac{\lambda_1+\lambda_2}{2}\right)-\frac{1}{2}\log(\lambda_1\lambda_2)$.
We check that $D_\CS(p_{\lambda_1},p_{\lambda_2})\geq 0$ since the arithmetic mean $A(\lambda_1,\lambda_2)=\frac{\lambda_1+\lambda_2}{2}$ is greater or equal than the geometric mean $G(\lambda_1,\lambda_2)=\sqrt{\lambda_1\lambda_2}$, and 
$D_\CS(p_{\lambda_1},p_{\lambda_2})=\log\frac{A(\lambda_1,\lambda_2)}{G(\lambda_1,\lambda_2)}$.

Choose $\omega=0$ so that $p_\lambda(\omega)=\lambda$ and $l_\lambda(\omega)=\log\lambda$.

\begin{eqnarray}
\rho(p_{\theta_1},p_{\theta_2}) &=&\frac{\sqrt{p_{2\theta_1}(\omega)}\sqrt{p_{2\theta_2}(\omega)}}{p_{\theta_1+\theta_2}(\omega)},\\
&=& \frac{2\sqrt{\lambda_1\lambda_2}}{\lambda_1+\lambda_2},\\
D_\CS(p_{\theta_1},p_{\theta_2})&=& l_{\theta_1+\theta_2}(\omega)-\frac{l_{2\theta_1}(\omega)+ l_{2\theta_2}(\omega)}{2},\\
&=& \log(\lambda_1+\lambda_2)-\frac{1}{2}(\log(2\lambda_1)+\log(2\lambda_2))=\log\left(\frac{\lambda_1+\lambda_2}{2\sqrt{\lambda_1\lambda_2}}\right).
\end{eqnarray}

To illustrate the fact that the formula Eq.~\ref{eq:rhoomega} is independent of the choice of $\omega$, let 
us consider $\omega=1$ so that $p_\lambda(\omega)=\lambda \exp(-\lambda)$ with log-likelihood $l_\lambda(\omega)=\log(\lambda)-\lambda$.
The correlation coefficient is then calculated as
\begin{eqnarray}
 \rho(p_{\theta_1},p_{\theta_2}) &=&\frac{\sqrt{p_{2\theta_1}(\omega)}\sqrt{p_{2\theta_2}(\omega)}}{p_{\theta_1+\theta_2}(\omega)},\\
&=& \frac{\sqrt{2\lambda_1\exp(-2\lambda_1)}  \sqrt{2\lambda_2\exp(-2\lambda_2)}}{(\lambda_1+\lambda_2)\exp(-(\lambda_1+\lambda_2))},\\
&=& \frac{2\sqrt{\lambda_1\lambda_2}}{\lambda_1+\lambda_2} \frac{\exp(-(\lambda_1+\lambda_2))}{\exp(-(\lambda_1+\lambda_2))},\\
&=& \frac{2\sqrt{\lambda_1\lambda_2}}{\lambda_1+\lambda_2}.
\end{eqnarray}

\subsection{Exponential family of Poisson distributions}

The Poisson family of probability mass functions (PMFs) 
$p_\lambda(x)=\frac{\lambda^x\exp(-\lambda)}{x!}$ where $\lambda>0$ denotes the intensity parameter and $x\in\calX=\{0,1,\ldots, \}$ is a discrete exponential family 
with sufficient statistic $t(x)=x$, natural parameter $\theta(\lambda)=\log \lambda$ (affine natural parameter space), cumulant function $F(\theta)=\exp(\theta)$, and auxiliary carrier term $k(x)=-\log x!$.
The informational energy is
\begin{eqnarray}
I(p_\lambda) &=&\sum_{x=0}^\infty p_\lambda^2(x),\\
 &=& I(p_\theta) =  \exp\left(F(2\theta)-2F(\theta)\right) E_{p_{2\theta}}\left[\exp(k(x))\right],\\
&=& \exp(e^{2\theta}-2e^{\theta}) E_{p_{2\theta}}\left[ \frac{1}{x!}\right],\\
&=& e^{\lambda^2-2\lambda} E_{p_{\lambda^2}}\left[ \frac{1}{x!}\right].
\end{eqnarray} 
 
\subsection{Exponential family of univariate normal distributions}

Consider the set of univariate normal probability density function:
\begin{equation}
\mathcal{N}:=\left\{ p_\lambda(x)=\frac{1}{\sigma \sqrt{2 \pi}} \exp\left(-\frac{1}{2}\left(\frac{x-\mu}{\sigma}\right)^{2}\right),\ \lambda=(\mu,\sigma^2)\in\bbR\times\bbR_{++}\right\}.
\end{equation}
Family $\calN$ is interpreted as an exponential family 
 indexed by the source parameter $\lambda=(\mu,\sigma^2)\in\Lambda$ with $\Lambda=\bbR\times\bbR_{++}$.
The corresponding natural parameter is $\theta(\lambda)=\left(\frac{\mu}{\sigma^2},-\frac{1}{2\sigma^2}\right)$
with the sufficient statistic $t(x)=(x,x^2)$ on the support $\calX=(-\infty,\infty)$ (and no additional carrier term, i.e., $k(x)=0$).
The cumulant function for the normal family is $F(\theta)=-\frac{\theta_{1}^{2}}{4 \theta_{2}}+\frac{1}{2} \log \left(-\frac{\pi}{\theta_{2}}\right)$.

We have
\begin{eqnarray}
I(p_\theta) &=& \exp\left(F(2\theta)-2F(\theta)\right),\\
&=& \exp\left(  - \frac{(2\theta_1)^2}{4(2\theta_2)}+\frac{1}{2}\log \frac{\pi}{(-2\theta_2)}+\frac{1}{2}\frac{\theta_1^2}{\theta_2}-\log\frac{\pi}{(-\theta_2)}  \right),\\
&=&  \exp\left( \frac{1}{2}\log \frac{\pi}{(-2\theta_2)}-\log\frac{\pi}{(-\theta_2)}   \right),\\
&=& \exp\left( -\frac{1}{2}\log(2\pi)-\frac{1}{2}\log(2\sigma^2)  \right),\\
&=& \frac{1}{2\sigma\sqrt{\pi}}.
\end{eqnarray}

Similar calculations for $\theta_1=(\mu_1,\sigma_1)$ and $\theta_2=(\mu_2,\sigma_2)$ yields
\begin{equation}
I(p_{\mu_1,\sigma_1},p_{\mu_2,\sigma_2}) = \frac{1}{\sqrt{2\pi}}  \frac{\exp\left(-\frac{(\mu_1-\mu_2)^2}{2\sigma_1^2+2\sigma_2^2}\right)}{\sqrt{\sigma_1^2+\sigma_2^2}}.
\end{equation}
We check that $I(p_\theta) = I(p_{\theta},p_{\theta})$.

It follows that Onicescu's correlation coefficient between two normal densities is:
\begin{equation}
\rho(p_{\mu_1,\sigma_1},p_{\mu_2,\sigma_2})= \sqrt{\frac{2\sigma_1\sigma_2}{\sigma_1^2+\sigma_2^2}} \exp\left(-\frac{(\mu_1-\mu_2)^2}{2\sigma_1^2+2\sigma_2^2}\right),
\end{equation}
and the  Cauchy-Schwarz divergence between two univariate Gaussians is:

\begin{equation}
D_\CS(p_{\mu_1,\sigma_1},p_{\mu_2,\sigma_2})=-\log\rho(p_{\mu_1,\sigma_1},p_{\mu_2,\sigma_2})= 
\frac{(\mu_1-\mu_2)^2}{2\sigma_1^2+2\sigma_2^2}+\frac{1}{2}\log \left(\frac{1}{2}\left(\frac{\sigma_1}{\sigma_2}+\frac{\sigma_2}{\sigma_1} \right)\right).\label{eq:csnormal1d}
\end{equation}

\subsection{Exponential family of multivariate normal distributions}

Consider the example of the multivariate normal (MVN) family:
The parameter $\lambda=(\lambda_v,\lambda_M)$ of a MVN consists of a vector part $\lambda_v=\mu$ and a $d\times d$ positive-definite matrix part $\lambda_M=\Sigma\succ 0$.
The density is given by
\begin{equation}
p_\lambda(x;\lambda) =  \frac{1}{(2\pi)^{\frac{d}{2}}\sqrt{|\lambda_M|}}  \exp\left(-\frac{1}{2} (x-\lambda_v)^\top \lambda_M^{-1} (x-\lambda_v)\right),
\end{equation}
where $|\cdot|$ denotes the matrix determinant.
Choose the sufficient statistic $t(x)=(x,-\frac{1}{2}xx^\top)$ so that $\theta=(\theta_v=\Sigma^{-1}\mu,\theta_M=\Sigma^{-1})$.
Since $k(x)=0$, let $\omega=0$, 
$$
p_\lambda(0)=\frac{1}{(2\pi)^{\frac{d}{2}}\sqrt{|\Sigma|}}\exp\left(-\frac{1}{2}\mu^\top\Sigma^{-1}\mu\right),
$$
and apply the formula of Eq.~\ref{eq:iecfk} with $2\theta_M=2\Sigma^{-1}=\left(\frac{1}{2}\Sigma\right)^{-1}$:
\begin{eqnarray}
I(p_\theta) &=& \frac{p_\theta^2(0)}{p_{2\theta}(0)},\\
&=& \frac{(2\pi)^{\frac{d}{2}}\left|\frac{1}{2}\Sigma\right|^{\frac{1}{2}}}{(2\pi)^d |\Sigma|}=\frac{1}{2^d \pi^{\frac{d}{2}} |\Sigma|^{\frac{1}{2}}}.
\end{eqnarray}

Let us calculate the formula for the Cauchy-Schwarz divergence between two multivariate Gaussian distributions.
We have $\theta(\lambda)=(\theta_b,\theta_M)=(\Sigma^{-1}\mu,\Sigma^{-1})$ for $\lambda=(\mu,\Sigma)$.
Conversely, we have $\lambda(\theta)=(\theta_M^{-1}\theta_v,\theta_M^{-1})$.
It follows that 
\begin{equation}
\lambda(\theta_1+\theta_2)=\left((\Sigma_1^{-1}+\Sigma_2^{-1})^{-1} (\Sigma_1^{-1}\mu_1+\Sigma_2^{-1}\mu_2)  , (\Sigma_1^{-1}+\Sigma_2^{-1})^{-1} \right).
\end{equation}
In particular, we have $\lambda(2\theta)=(\mu,\frac{1}{2}\Sigma)$.
Let $\omega=0$ so that
\begin{eqnarray}
p_\lambda(0) &=& \frac{1}{(2\pi)^{\frac{d}{2}}\sqrt{|\Sigma|}}\exp\left(-\frac{1}{2}\mu^\top\Sigma^{-1}\mu\right),\\
l_\lambda(0) &=& -\frac{d}{2}\log(2\pi)-\frac{1}{2}\log|\Sigma|-\frac{1}{2}\mu^\top\Sigma^{-1}\mu.
\end{eqnarray}

Thus we get

\begin{eqnarray}
D_\CS(p_{\lambda_1},p_{\lambda_2}) &=& l_{\lambda(\theta_1+\theta_2)}(\omega)-\frac{l_{\lambda(2\theta_1)}(\omega)+ l_{\lambda{2\theta_2}}(\omega)}{2},\\
D_\CS(p_{\mu_1,\Sigma_1},p_{\mu_2,\Sigma_2})  &=& \frac{1}{2}\log\left(\frac{1}{2^d} \frac{\sqrt{|\Sigma_1|\ \Sigma_2||}}{|(\Sigma_1^{-1}+\Sigma_2^{-1})^{-1}|} \right)\nonumber\\
&& +\frac{1}{2}\mu_1^\top\Sigma_1^{-1}\mu_1+\frac{1}{2}\mu_2^\top\Sigma_2^{-1}\mu_2\nonumber\\
&& -\frac{1}{2}(\Sigma_1^{-1}\mu_1+\Sigma_2^{-1}\mu_2)^\top (\Sigma_1^{-1}+\Sigma_2^{-1})^{-1}  (\Sigma_1^{-1}\mu_1+\Sigma_2^{-1}\mu_2).
\end{eqnarray}

The formula coincides with the formula of Eq.~\ref{eq:csnormal1d} when $\Sigma_1=\sigma_1^2$ and $\Sigma_2=\sigma_2^2$.

\subsection{Exponential family of Pareto distributions}

Consider the family of Pareto densities defined by a shape parameter $a>0$ and a prescribed scale parameter $k>0$ as follows:
\begin{equation}
\left\{p_{a}(x)=\frac{ak^a}{x^{a+1}},\quad x\in [k,\infty)\right\}.
\end{equation}
Writing the density as $p_{a}(x)=\exp(a\log k+\log a-(a+1)\log x)=p_{\theta(a)}(x)$, we deduce that
the Pareto densities form an exponential family with natural parameter $\theta=a+1$,   sufficient statistic $t(x)=\log x$ and $k(x)=0$. 
Let us choose $\omega=k$, and apply the generic formula for the informational energy with $\theta(a)=a+1$ and $\theta^{-1}(b)=b-1$ (and $2\theta(a)=\theta^{-1}(2a+2)=2a+2-1=2a+1$):
\begin{eqnarray}
I(p_\theta) &=&  \frac{p_\theta^2(\omega)}{p_{2\theta}(\omega)},\\
I(p_a)&=&  \frac{p_{\theta(a)}^2(k)}{p_{2\theta(a)}(k)}  =  \frac{p_{a}^2(k)}{p_{2a+1}(k)}\\
&=& \left(\frac{ak^a}{x^{a+1}}\right)^2 \frac{k^{2a+2}}{(2a+1)k^{2a+1}}, \\ 
&=& \frac{a^2}{k(2a+1)}.
\end{eqnarray}

\subsection{Instantiating formula with a computer algebra system}

In general, we can automate  the calculations of closed-form formula for conic exponential families using a computer algebra system  (CAS)  by defining the source-to-natural parameter conversion function $\theta(\lambda)$, and then apply the formula
\begin{equation}
D_\CS(p_{\lambda_1},p_{\lambda_2})= \log\left(  \frac{p_{\theta(\lambda_1)+\theta(\lambda_2)}(\omega)}{\sqrt{p_{2\theta(\lambda_1)}(\omega)}\sqrt{p_{2\theta(\lambda_2)}(\omega)}} \right),\quad \forall \omega\in\calX.
\end{equation}

For example, using the CAS  {\tt Maxima} (\url{http://maxima.sourceforge.net/}), we can calculate the formula of the information energy of Pareto densities as follows:

\begin{verbatim}
/* Pareto densities form an exponential family */
assume(k>0);
assume(a>0);

Pareto(x,a):=a*(k**a)/(x**(a+1));

/* check that it is a density (=1) */
integrate(Pareto(x,a),x,k,inf);

/* calculate Onicescu's informational energy */
integrate(Pareto(x,a)**2,x,k,inf);

/* method bypassing the integral calculation */
omega:k;
(Pareto(omega,a)**2)/Pareto(omega,2*a+1);
\end{verbatim}

\bibliographystyle{plain}
\bibliography{OnicescuBIB}

\end{document}